\documentclass[11pt,a4paper]{article}
\pdfoutput=1
\usepackage{jheppub}

\usepackage[utf8]{inputenc}
\usepackage{graphicx}
\usepackage{amsmath,amsfonts,amssymb}
\usepackage{epsfig,color}
\usepackage[outdir=./]{epstopdf}
\usepackage[T1]{fontenc}
\usepackage{physics}
\usepackage{graphicx}
\usepackage{color} 
\usepackage{dcolumn}
\usepackage{bm}

\usepackage[subfigure]{graphfig}
\usepackage{latexsym}
\usepackage{amsmath}
\usepackage{amsfonts}
\usepackage{amsxtra}
\usepackage{hyperref}
\usepackage{placeins}

\usepackage[normalem]{ulem}

\usepackage{slashed}
\usepackage{graphicx}
\usepackage{mathrsfs}

\usepackage{mathtools}
\usepackage{simpler-wick}
\allowdisplaybreaks[4]

\newcommand{\msbar}{$\overline{\mbox{{\rm MS}}} \ $}

\newcommand{\lwrsim}{\raise0.3ex\hbox{$<$\kern-0.75em\raise-1.1ex\hbox{$\sim$}}}
\def\krto{ {\,\,\lower .8ex\hbox {$\longrightarrow \atop k \rightarrow 0$}\,\,}}

\def\bea{\begin{eqnarray} }
\def\beq{\begin{eqnarray} }

\def\eea{\end{eqnarray}}
\def\eeq{\end{eqnarray}}

\newcommand{\nn}{\nonumber \\ }

\begin{document}


\title{Parton distribution function  for  topological charge at one loop}

\author{Anatoly Radyushkin and Shuai Zhao}

\affiliation{Old Dominion University,  \\4600 Elkhorn Ave., Norfolk, VA 23529, USA}
\affiliation{Thomas Jefferson National Accelerator Facility,  \\ 12000 Jefferson Ave., Newport News, VA 23606, USA}

\emailAdd{radyush@jlab.org}
\emailAdd{szhao@odu.edu}

\abstract{
We present  results for the   $gg$-part of the  one-loop corrections to 
the recently introduced ``topological charge'' GPD $\widetilde F(x,q^2)$.
In particular, we give expression for    its evolution kernel. 
To enforce  strict compliance with the gauge invariance requirements,
we have used  on-shell states for external gluons, and have obtained identical results 
both in Feynman and light-cone gauges. 
No ``zero mode'' 
 $\delta (x)$  terms were found for the  twist-4 gluon GPD  $\widetilde  F(x,q^2)$.

}
  
\maketitle

\date{\today}

\flushbottom




\section{
 Introduction}

Parton distribution functions (PDFs) $f(x)$ \cite{Feynman:1973xc}  
 provide an efficient way to describe  hadron structure. At present,
  PDFs are  the objects of both  intensive experimental research and lattice QCD calculations.
  In fact, it is believed that  the lattice studies may provide information about interesting  PDFs that 
  are difficult or impossible to  investigate in accelerator experiments. 
  Among such parton distributions, one may list two twist-4 gluon functions proposed recently in Refs. 
  \cite{Hatta:2020iin,Hatta:2020ltd}.  
  
   One of them,  introduced in Ref.  \cite{Hatta:2020iin}  and denoted there as  $F(x)$,  describes the 
  momentum distribution of  the 
``gluon condensate''.   
It  corresponds to the forward matrix element of the  bilocal operator  $\langle p | F_{\mu \nu} (0) W[0,z]F^{\mu \nu} (z)|p \rangle$
taken on the light cone $z=z^-$. 
 The $x$-integral of $F(x)$ corresponds to the 
 matrix element    $\langle p | F_{\mu \nu} (0) F^{\mu \nu} (0)|p \rangle$  of the local operator that  
 may be related to the gluon contribution into the proton mass. 
The  second example  is the twist-4 gluon distribution discussed in Ref.  \cite{Hatta:2020ltd}. 
It is defined through 
   the bilocal  operator  $F_{\mu \nu} (0) W[0,z] \widetilde F^{\mu \nu} (z)$ corresponding   
to  the ``topological charge'' density. 
Since the forward matrix element of this operator between the nucleon states vanishes,
it was proposed in \mbox{Ref. \cite{Hatta:2020ltd} }to consider the   non-forward matrix element
 $\langle p' | F_{\mu \nu} (0) W[0,z] F^{\mu \nu} (z)|p \rangle$, 
i.e.,  the relevant generalized parton distribution function (GPD).  
The simplest situation  corresponds to ``zero skewness'',  when the momentum 
transfer $q \equiv p'-p$ satisfies $q\cdot z=0$, and one deals with the function 
$\widetilde F(x, q^2)$ of the light-cone  momentum  fraction $x$ and the momentum \mbox{transfer $q^2$.}

 A rather intriguing  question raised in Ref.  \cite{Ji:2020baz}  is whether
 twist-4 gluon PDFs have   singular $\delta (x)$ ``zero-mode'' contributions, 
 similar to those that  have been found  \cite{Burkardt:2001iy}
 in calculations of one-loop perturbative QCD corrections for the 
 twist-3 quark PDFs.     For  $F(x)$, this question   was  originally   investigated in Ref. \cite{Hatta:2020iin}.
 However, 
  the matrix element of the bilocal operator 
$F_{\mu \nu} (z) F^{\mu \nu} (0)$   in the  calculation of  Ref. \cite{Hatta:2020iin} was taken  between  gluon states
with nonzero virtuality. This is a risky exercise because it violates gauge invariance. 
Indeed, as shown in our paper  \cite{Radyushkin:2021fel}, the calculations with virtual external gluon lines 
in Feynman and light-cone gauges give 
different results, both of which are incorrect.

To  perform the  calculation in a gauge-invariant way, 
 one needs  to  do the   calculations  
using  on-shell external gluons. 
However,  there is a complication that both the    tree-level  and one-loop 
matrix elements  of the $F_{\mu \nu} (0) F^{\mu \nu} (z)$ operator 
 for  on-shell gluon  states  vanish.  
 To escape  this problem, we   took   a nonforward matrix element, i.e. considered 
the  generalized parton distribution (GPD) corresponding to the same bilocal operator $F_{\mu \nu} (0) F^{\mu \nu} (z)$.

 In the  ``topological charge''  case,  the forward matrix element of 
 $F_{\mu \nu} (0) \widetilde F^{\mu \nu} (z)$  operator
 vanishes, even  if  the external gluons are  off-shell. 
 Hence, the use of a nonforward kinematics is mandatory. 
 The calculation of  the relevant GPD at one-loop level was done in Ref.   \cite{Hatta:2020ltd}, but 
still using  off-shell gluons.  

Our goal in the present paper is to perform a one-loop calculation 
for  the nonforward matrix element of the  $F_{\mu \nu} (0) \widetilde F^{\mu \nu} (z)$
operator between on-shell gluon states. We give a rather detailed description of our calculations,
displaying intermediate diagram-by-diagram results 
 both in Feynman and light-cone gauges. 
 We also list all one-loop integrals    necessary 
 to get these   results. The total result   is the same in both gauges. 
However, it  is different from the result given in Ref.  \cite{Hatta:2020ltd}.

The content of the paper is organized as follows. 
In Section 2,   we discuss the definition of the GPD $\widetilde F(x,q^2)$ 
 related  to a nonforward 
matrix element involving  on-shell gluons. In Section 3, we present diagram-by-diagram  results for 
all contributing one-loop 
 diagrams in Feynman gauge.  In Section 4, we discuss the results of calculations
 made in the light-cone gauge. In Section 5,  we write down  the total result and discuss its structure. 
In \mbox{Section 6,}    we give a summary of  the paper and  discuss further steps in the study of twist-4 gluon distributions.
 The table of basic integrals that appear in our calculations is given in the Appendix.

\section{Parton distribution   for topological charge}

The gluon GPD   $\widetilde F(x,q^2)$ corresponding to the momentum  distribution of the topological charge is defined
 \cite{Hatta:2020ltd}  through a nonforward matrix element of the twist-4 
bilocal combination of the gluon fields
\begin{align}
	\widetilde F(x,q^2)=P^+\int_{-1}^1  \frac{\dd z^{-}}{2 \pi} e^{i x P^{+} z^{-}}\left\langle p'\left|F^{\mu \nu}(0) W[0, z^-] \widetilde{F}_{\mu \nu}\left(z^{-}\right)\right| p\right\rangle \ 
	\label{tFdef}
\end{align}
switched between the nucleon states having  momenta $p,p'$, with  $P=(p+p')/2$ being the average momentum 
and $q=p'-p$ the momentum transfer. 
As usual,  $\widetilde{F}_{\mu\nu}=\frac12\varepsilon_{\mu\nu\alpha\beta}F^{\alpha\beta}$, and $\varepsilon_{\mu\nu\alpha\beta}$ is the Levi-Civita tensor.
The summation over the gluon colors  and division by their  number  $N_g=N_c^2-1$  is assumed.  Also, 
 the  summation over the  hadron polarizations is implied. 
The gluon fields $F(0)$ and $ \widetilde{F}(z_-)$  are connected by  the straight-line gauge link $W[0, z^-]$  in the ``minus''  direction
specified by  the  light-cone vector $n$.    
The ``plus''-components  for an arbitrary vector $a$ are obtained by a scalar product with  $n$, i.e.,   $a^{+} =n \cdot a\ .$

In    QCD,  PDFs and GPDs   also have  a dependence on the factorization scale $\mu$. The latter  emerges 
as an ultraviolet cut-off in  the   perturbative corrections to the relevant operator on the light cone. 
To calculate such corrections in the momentum representation, one needs to consider the matrix element (\ref{tFdef})
between the parton states. In the present paper,  we will study  the case of gluon external states $ | g(p,\epsilon) \rangle $,
where $p$ is the gluon momentum and $\epsilon$ is its polarization. 
It is instructive to consider first  the tree level expression  for  the   
forward matrix element. We have 
\begin{align}
 &p^+\int \frac{\dd z^{-}}{2 \pi} e^{i x p^{+} z^{-}}\left\langle g(p,\epsilon_2^*)\left | F^{\mu \nu}(0) W[0, z] \widetilde{F}_{\mu \nu}\left(z^{-}\right)\right| g(p,\epsilon_1)\right\rangle^{(0)}\nonumber\\
 =&{\frac12} n\cdot p \, (p^{\mu}\epsilon_2^{*\nu}-p^{\nu}\epsilon_2^{*\mu}) \varepsilon_{\alpha\beta\mu\nu}
 (p^{\alpha}\epsilon_1^{\beta}-p^{\beta}\epsilon_1^{\alpha}) 
 \Big [ \delta(n\cdot p-x\,  n\cdot p)+ \delta(n\cdot p+x\,  n\cdot p) \Big ]
 \nn & = \, {2} \varepsilon_{\alpha\beta\mu\nu} p^{\mu}\epsilon_2^{*\nu} p^{\alpha}\epsilon_1^{\beta} \Big [ \delta (1-x)+ \delta (1+x) \Big ] =0
 \  .
\end{align}
We took here different gluon polarizations $\epsilon_1$ and $\epsilon_2$  for the initial and final states. Still, 
the tree-level matrix element vanishes because the momentum vector $p$ enters twice in the convolution with the Levi-Civita tensor.
Moreover,   this happens  no matter if  the gluons are  on-shell or not.
To  get a nonzero result in the  $\varepsilon_{\alpha\beta\mu\nu} \ldots$ convolution, we need  another vector instead of one of the ``$p$'' factors. 
To this end,  we shall consider (just like in Ref.  \cite{Hatta:2020ltd})  the function defined by a  non-forward matrix element
\begin{align}
F(x,\xi,q^2)=&\frac{P^+}{N_g}\int \frac{\dd z^{-}}{2 \pi} e^{i x P^{+} z^{-}}
\nn &
\left\langle g(p+q,\epsilon_2^*)\left | 
F^{a, \mu \nu} \left (-\frac{z^-}{2} \right ) 
W \left [-\frac{z}{2}, \frac{z}{2} \right ] \widetilde F_{a,\mu \nu}\left(\frac{z^{-}}{2}\right)\right| g(p,\epsilon_1)\right\rangle,
\end{align}
where $P=\frac{p+(p+q)}{2}$.  In general, the skewness is  defined  by  $\xi\equiv -\frac{q^+}{2P^+}$, so  that  $n\cdot p=(1+\xi)n\cdot P$. 
However, in  the present work, we take   $\xi=0$. 
 Furthermore, we use the gluons  that   (unlike Ref.  \cite{Hatta:2020ltd}) are on-shell both  in the  initial and final states , i.e.,
\begin{align}
p^2=0,~~(p+q)^2=0, ~~p\cdot \epsilon_1=0,~~(p+q)\cdot \epsilon_2^*=0 \ .
\end{align}
 It is convenient to    take  also {$n\cdot \epsilon_1=n\cdot \epsilon_2=0$}.
The tree-level result is now given by 
\begin{align}
F^{(0)}(x,q^2)=&\frac12 n\cdot P \left ((p+q)^{\mu}\epsilon_2^{*\nu}-(p+q)^{\nu}\epsilon_2^{*\mu} \right )\,  \varepsilon_{\alpha\beta \mu\nu } \, 
(p^{\alpha}\epsilon_1^{\beta}-p^{\beta}\epsilon_1^{\alpha})
\nn & \times   \Big [ \delta(n\cdot P-x\,  n\cdot P)+ \delta(n\cdot P+x\,  n\cdot P) \Big ] \nonumber\\
=&\frac12 \varepsilon_{\alpha\beta \mu\nu }(p^{\alpha}\epsilon_1^{\beta}-p^{\beta}\epsilon_1^{\alpha})(q^{\mu}\epsilon_2^{*\nu}-q^{\nu}\epsilon_2^{*\mu})
 \Big [ \delta (1-x)+ \delta (1+x) \Big ] \nonumber\\
=&- 2 \varepsilon(p,q,\epsilon_1,\epsilon_2^*) \Big [ \delta (1-x)+ \delta (1+x) \Big ] \nonumber\\
\equiv &\Pi(p,q,\epsilon_1,\epsilon_2^*)  \Big [ \delta (1-x)+ \delta (1+x) \Big ] \  , 
\label{F0}
\end{align}
where we have denoted $F(x,\xi=0,q^2)=F(x,q^2)$,  
\begin{align}
	\varepsilon(p,q,r,s)\equiv \varepsilon^{\alpha\beta\gamma\delta}p_{\alpha}q_{\beta}r_{\gamma}s_{\delta} \ , 
\end{align}
and
\begin{align}
\Pi(p,q,\epsilon_1,\epsilon_2^*)=-2 \varepsilon(p,q,\epsilon_1,\epsilon_2^*).
\end{align}

\section{One-loop corrections {in Feynman gauge}}

Our goal is to investigate the structure of  this matrix element at the one-loop level. 
 To be on safe side, we have performed  our calculations  
 both in the  light-cone gauge and in Feynman gauge.  The  gluon propagator  in the light-cone gauge  is given by $-i D^{\mu\nu}(k)/k^2$, where
\begin{align}
D^{\mu\nu}(k)=g^{\mu\nu}-\frac{k^{\mu}n^{\nu}+k^{\nu}n^{\mu}}{n\cdot k}.
\end{align}
In Feynman gauge, we have 
\begin{align}
D^{\mu\nu}(k)=g^{\mu\nu}.
\end{align}
To handle ultraviolet and collinear divergences, we use 
the  dimensional regularization, defining  the dimension $d$ of space-time by $d=4-2\epsilon$.

\begin{figure}[t]
\centerline{\includegraphics[width=5in]{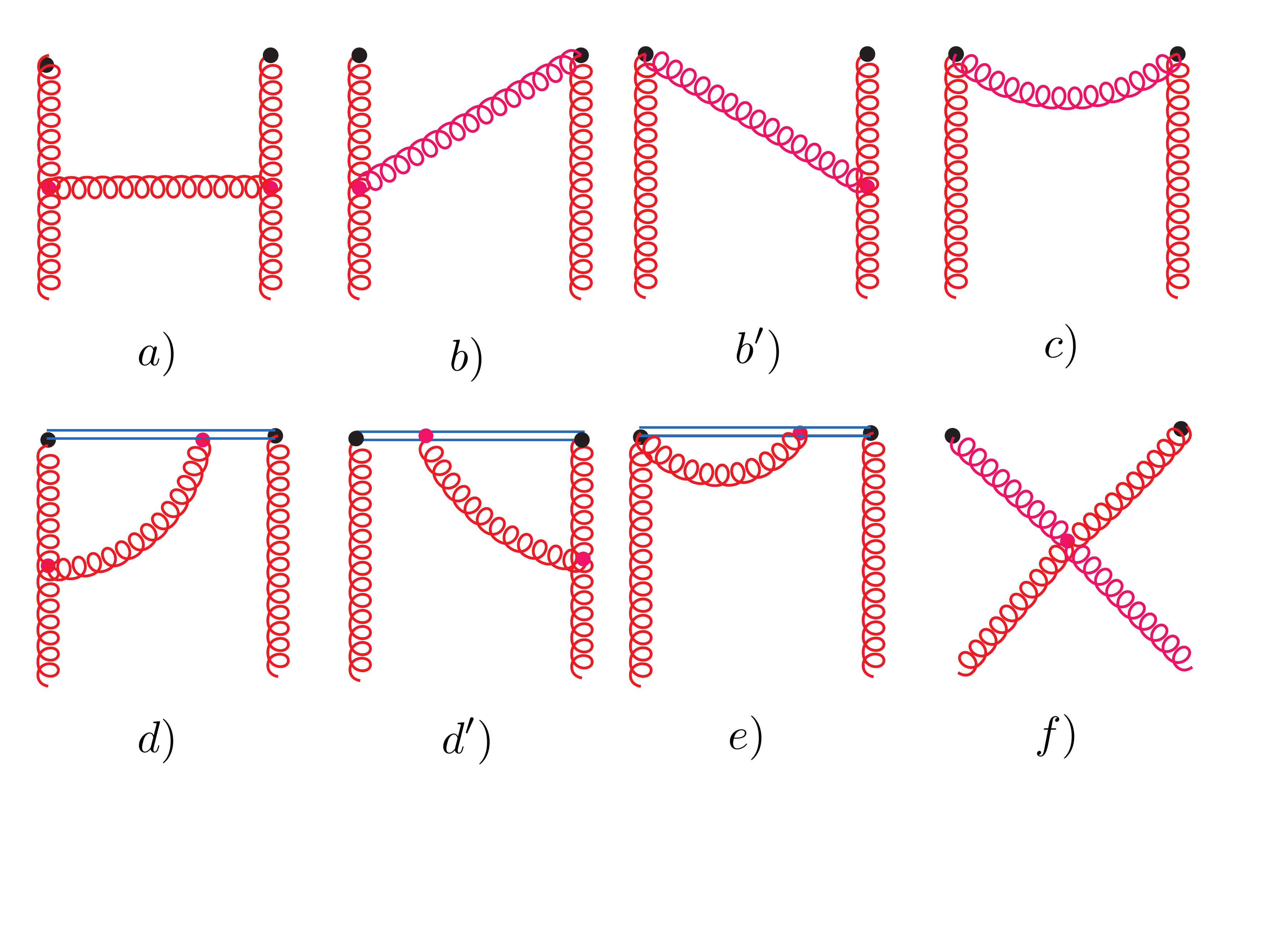}} \vspace{-20mm}
\centerline{\includegraphics[width=5.2in]{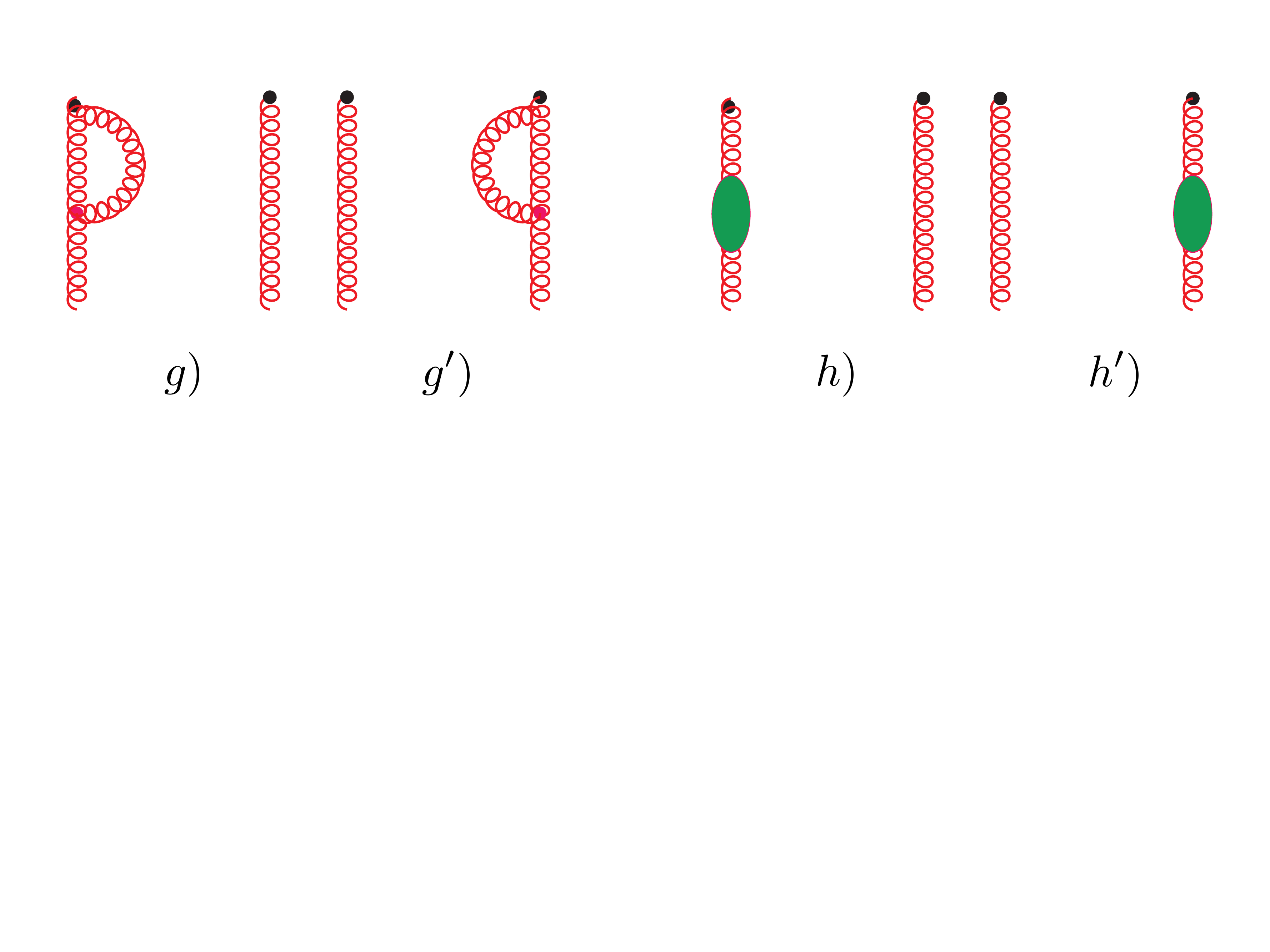}} 
\vspace{-55mm}
 \caption{ One-loop  diagrams  (mirror diagram $e'$  is not shown). 
  \label{real}}
\end{figure}

In this section, we discuss    calculations in  Feynman gauge. The 
relevant   diagrams are shown in Fig.~\ref{real}.
We will express the results for particular diagrams in terms of basic integrals
	\begin{align}
	&S_{lmn}
	=\int\frac{\dd ^dk}{(2\pi)^d}\delta \left (x-\frac{n\cdot k}{n\cdot P} \right )\frac{1}{D_1^l  D_2^m D_3^n} \ , 
	\end{align}
\begin{align}
	&V^{\mu}_{lmn}
	=\int\frac{\dd ^dk}{(2\pi)^d}\delta  \left  (x-\frac{n\cdot k}{n\cdot P}  \right  )\frac{k^{\mu}}{D_1^l  D_2^m D_3^n} \ , 
	\end{align}
	\begin{align}
	&T^{\mu \nu}_{lmn}
	=\int\frac{\dd ^dk}{(2\pi)^d}\delta  \left  (x-\frac{n\cdot k}{n\cdot P}  \right  )\frac{k^{\mu}k^{\nu}}{D_1^l  D_2^m D_3^n} \ ,
	\end{align} 	where $D_1=k^2$, $D_2 = (p-k)^2$, $D_3 = (k+q)^2$. 

\subsection{Box diagram}

For the ``box''  diagram shown in  Fig.~(1a), we have
\begin{align}
\widetilde F_{(1a)}&  \left(x, p,q, \epsilon_1, \epsilon_2\right) \Big |_{x\geq 0} 	={g^2 C_A} \cdot 2i\varepsilon_{\alpha\mu\nu\rho}\bigg[({V_{011}^{\alpha}}-V_{101}^{\alpha}+V_{110}^{\alpha}-2q^2 V_{111}^{\alpha})q^{\mu}\epsilon_{1}^\nu\epsilon_{2}^{*\rho} \nonumber\\
	&+ 4 p^{\mu}q^{\nu}\epsilon_1^{\rho}(-2T_{111}^{\alpha\beta}\epsilon_{2,\beta}^*+V_{111}^{\alpha}p\cdot\epsilon_2^*)+4 p^{\mu}q^{\nu}\epsilon_2^{*\rho}(2T_{111}^{\alpha\beta}\epsilon_{1,\beta}+V_{111}^{\alpha}q\cdot \epsilon_1)\bigg]
\end{align}	
	Using explicit expressions for the basic integrals (see the Appendix)  and simplifying, we obtain 
\begin{align}
	 \widetilde F_{(1a)}&  \left(x, p,q, \epsilon_1, \epsilon_2\right)  \Big |_{x\geq 0} 	= -\delta(x)\frac{g^2 C_A}{8\pi^2 {n\cdot P}} 
	 \left (\frac{\mu^2 e^{\gamma_E}}{-q^2} \right )^{\epsilon}
	  \frac{\Gamma^2(1-\epsilon)\Gamma(\epsilon)}{
	   \Gamma(2-2\epsilon)}
	 q^2 \varepsilon(n,q,\epsilon_1,\epsilon_2^*)\nonumber\\
	&-\delta'(x)\frac{g^2 C_A q^2}{16\pi^2 {n\cdot P}}\left (\frac{\mu^2 e^{\gamma_E}}{-q^2} \right )^{\epsilon}
	\frac{\Gamma^2(2-\epsilon)\Gamma(-1+\epsilon)}{ \Gamma(4-2\epsilon)}\varepsilon(n,q,\epsilon_1,\epsilon_2^*)\nonumber\\
	&+{\theta(0\leq x\leq 1)}\frac{g^2 C_A}{8\pi^2 {n\cdot P} }\bigg[2\, n\cdot p \,  \varepsilon(p,q,\epsilon_1,\epsilon_2^*)\bigg(-\frac{4 (1-x)^{1-2\epsilon}\Gamma(1-\epsilon)^2 \Gamma(\epsilon)}{\Gamma(2-2\epsilon)}\left (\frac{\mu^2 e^{\gamma_E}}{-q^2} \right )^{\epsilon}\nonumber\\
	&+x \left (-\frac{1}{\epsilon}+\frac{1}{\epsilon_{\mathrm{IR}}}-\ln\frac{\mu^2}{\mu_{\mathrm{IR}}^2}+{\bigg(\frac{\mu_{\mathrm{IR}}^2 e^{\gamma_E}}{-q^2}\bigg)^{\epsilon_{\mathrm{IR}}}}\frac{(1-x)^{-1-2\epsilon_{\mathrm{IR}}}\Gamma(-\epsilon_{\mathrm{IR}})^2 \Gamma(1+\epsilon_{\mathrm{IR}})}{\Gamma(-2\epsilon_{\mathrm{IR}})} \right )\bigg)\nonumber\\
	&+\frac{(1-x)^{-2\epsilon}\Gamma^2(1-\epsilon)\Gamma(\epsilon)}{\Gamma(2-2\epsilon)} \left (\frac{\mu^2 e^{\gamma_E}}{-q^2} \right )^{\epsilon}\bigg( q^2 (1-2\epsilon)\varepsilon(n,q,\epsilon_1,\epsilon_2^*)\nonumber\\
	&+2 \big [\varepsilon(n,p,q,\epsilon_1)q\cdot \epsilon_2^* +\varepsilon(n,p,q,\epsilon_2^*)q\cdot \epsilon_1 \big ]
	 \Big (1-2x (1-\epsilon) 
	\Big )
	\bigg)
	\bigg] \ . 
\end{align}

Note that, in addition to the   $\varepsilon (p, q, \epsilon_{1}, \epsilon_{2})$ structure, there are other ones. Let us show that the other structures can be reduced to 
$\varepsilon (p, q, \epsilon_{1}, \epsilon_{2})$. 
 Indeed, if the vectors $p, q, \epsilon_1, \epsilon_2$ are linearly independent in the $d=4$ space-time, the vector $n$ can be expressed in terms of the 
 other 4  vectors as
\begin{align}
	n^{\mu}=a_1 p^{\mu}+a_2 (p+q)^{\mu}+a_3 \epsilon_1^{\mu}+a_4\epsilon_2^{\mu} \ . 
\end{align}
Contracting above equation with $p$, $p+q$ and $n$ respectively,  we have
\begin{align}
	n\cdot p&=-a_2 \frac{q^2}{2} -a_4 q\cdot \epsilon_2,\\
	n\cdot p+n\cdot q&=-a_1 \frac{q^2}{2}+a_3 q\cdot \epsilon_1,\\
	0&=a_1 n\cdot p+a_2 (n\cdot p+n\cdot q).
\end{align}
In the zero-skewness  case, we have $n\cdot q = 0$, hence $a_1+a_2=0$ and  also
	\begin{align}\label{eq:str1}
	\varepsilon(n,p,q,\epsilon_1)q\cdot \epsilon_2^* +\varepsilon(n,p,q,\epsilon_2^*)q\cdot \epsilon_1
	=2 \, n\cdot p \,   \varepsilon(p,q,\epsilon_1,\epsilon_2^*)
\end{align}
Similarly, for   the $\varepsilon(n,q,\epsilon_1,\epsilon_2^*)$ structure, we have 
\begin{align}\label{eq:str2}
  \varepsilon(n,q,\epsilon_1,\epsilon_2^*)	=(a_1+a_2)  \varepsilon(p,q,\epsilon_1,\epsilon_2^*) =0
\end{align}
As a result, the terms proportional to $\delta (x)$ and $\delta'(x)$ vanish, and the remaining terms may be  written as 
\begin{align}
 \widetilde F_{(1a)}&  \left(x, p,q, \epsilon_1, \epsilon_2\right)  \Big |_{x\geq 0} ={\theta(0\leq x \leq 1)}\frac{g^2 C_A}{4\pi^2} \varepsilon(p,q,\epsilon_1,\epsilon_2^*)\nn & \times \bigg[x \left (-\frac{1}{\epsilon}+\frac{1}{\epsilon_{\mathrm{IR}}}-\ln\frac{\mu^2}{\mu_{\mathrm{IR}}^2}+\left (\frac{\mu_{\mathrm{IR}}^2 e^{\gamma_E}}{-q^2}\right )^{\epsilon_{\mathrm{IR}}}\frac{(1-x)^{-1-2\epsilon_{\mathrm{IR}}}\Gamma^2(-\epsilon_{\mathrm{IR}}) \Gamma(1+\epsilon_{\mathrm{IR}})}{\Gamma(-2\epsilon_{\mathrm{IR}})} \right )
 \nonumber\\
&-\frac{2(1-x)^{-2\epsilon}\Gamma^2(1-\epsilon)\Gamma(\epsilon)}{\Gamma(2-2\epsilon)}\left (\frac{\mu^2 e^{\gamma_E}}{-q^2} \right )^{\epsilon}  (1-2x\epsilon)
\bigg].
\end{align}
Thus, the  box diagram has both ultraviolet (UV) and infrared (IR) singular contributions,
reflected by the UV poles $1/\epsilon$ and IR poles $1/\epsilon_{\mathrm{IR}}$.

\subsection{Bremsstrahlung diagrams} 

For the diagram (1d),  containing an  insertion into the gluon link, we have
\begin{align}
	&{g^2 C_A}\frac{2i}{n\cdot P (1-x)}  \varepsilon_{\alpha\mu\nu\rho}(p+q)^{\mu}\epsilon_2^{*\rho} \bigg[n\cdot P (1+x) \epsilon_1^{\nu} V_{110}^{\alpha}-2  n^{\nu}  \epsilon_{1\beta} T_{110}^{\alpha\beta}\bigg]\nonumber\\
=&-\frac{\alpha_s C_A}{2\pi}\varepsilon(p,q,\epsilon_1,\epsilon_2^*)\left (\frac{1}{\epsilon}-\frac{1}{\epsilon_{\mathrm{IR}}}+\ln\frac{\mu^2}{\mu_{\mathrm{IR}}^2} \right )\bigg[\frac{x(1+x)}{1-x}{\theta(0\leq x\leq 1)}\bigg]_+  \  . 
\end{align}
For the mirror diagram (${\rm 1d}'$), we obtain 
\begin{align}
	&- {g^2 C_A}\frac{2i}{n\cdot P (1-x)}
	\varepsilon_{\alpha\mu\nu\rho}p^{\mu}\epsilon_1^{\nu}
	\bigg[n\cdot P (1+x) \epsilon_2^{*\rho}      
	\left (V_{011}^{\alpha}	+ q^{\alpha} S_{011} \right ) 
	\nonumber\\
	&- 2 n^{\rho}
	\left (\epsilon_{2,\beta}^* \left (T_{011}^{\alpha\beta} + q^{\alpha} V_{011}^{\beta}  \right  ) -
	p\cdot \epsilon_2^* 	\left (V_{011}^{\alpha}	+ q^{\alpha} S_{011} \right ) 
		 \right )  
	\bigg]\nonumber\\
=&-\frac{\alpha_s C_A}{2\pi}\varepsilon(p,q,\epsilon_1,\epsilon_2^*) \left (\frac{1}{\epsilon}
-\frac{1}{\epsilon_{\mathrm{IR}}}+\ln\frac{\mu^2}{\mu_{\mathrm{IR}}^2} \right )\bigg[\frac{x(1+x)}{1-x}{\theta(0\leq x\leq 1)}\bigg]_+ \ .
\end{align}
Thus, the final expressions for contributions of  diagrams (1d) and (${\rm 1d}'$)  coincide, and their combined contribution is given by 
\begin{align}
 \widetilde F_{(1d)+(1d')}  \left(x, p,q, \epsilon_1, \epsilon_2\right)  \Big |_{x\geq 0}	
=&-\frac{\alpha_s C_A}{\pi}\varepsilon(p,q,\epsilon_1,\epsilon_2^*) \left (\frac{1}{\epsilon}
-\frac{1}{\epsilon_{\mathrm{IR}}}+\ln\frac{\mu^2}{\mu_{\mathrm{IR}}^2} \right )
\nn & \times \bigg[\frac{x(1+x)}{1-x}{\theta(0\leq x\leq 1)}\bigg]_+ \ .
\label{brems}
\end{align}

Note, that 
these  diagrams  contain  the $\sim 1/(1-x)$ ``bremsstrahlung''  or  soft-gluon exchange  term. 
The  singularity for $x=1$ comes here   regularized by the ``plus'' prescription.
In fact, the diagram (1a) also has the $\sim 1/(1-x)$ contribution, but it is  not 
accompanied by the ``plus'' prescription. Namely,   it comes from the term containing
$(1-x)^{-1-2\epsilon_{\mathrm{IR}}}$. 

To combine the contributions of the diagrams 
(1a), (1d) and  (${\rm 1d}'$), we write the expression for the  diagram (1a) 
as a sum of a term having the plus-prescription for $x=1$  and a $\delta (1-x)$ term.
Expanding in $\epsilon$, ${\epsilon_{\mathrm{IR}}}$, and neglecting the terms vanishing 
when $\epsilon = 0$, ${\epsilon_{\mathrm{IR}}} = 0$, we obtain 
\begin{align}
 \widetilde F_{(1a)}&  \left(x, p,q, \epsilon_1, \epsilon_2\right)  \Big |_{x\geq 0} = 
\frac{\alpha_s}{\pi }C_A \epsilon(p,q,\epsilon_1,\epsilon_2^*)\bigg\{{\theta(0\leq x\leq 1)}
\nn \times &
\bigg[-\frac{2+x}{\epsilon}+\frac{x(1+x)}{1-x} 
\left (- \frac{1}{\epsilon_{\mathrm{IR}}}+\ln\frac{\mu^2}{\mu_{\mathrm{IR}}^2} \right )
-\frac{2}{1-x}\ln\frac{\mu^2}{-q^2(1-x)^2}-4(1-x)\bigg]\bigg\}_+\nonumber\\
&+\frac{\alpha_s}{\pi}C_A \epsilon(p,q,\epsilon_1,\epsilon_2^*)
 \delta(1-x)   \nn  & \times \bigg[\frac{1}{\epsilon_{\mathrm{IR}}^2}+\frac{1}{\epsilon_{\mathrm{IR}}}\ln\frac{\mu_{\mathrm{IR}}^2}{-q^2}+\frac12 \ln^2\frac{\mu_{\mathrm{IR}}^2}{-q^2}-\frac{\pi^2}{12}-2-\frac{5}{2}\bigg(\frac{1}{\epsilon}-\frac{1}{\epsilon_{\mathrm{IR}}}+\ln\frac{\mu^2}{\mu_{\mathrm{IR}}^2}\bigg)\bigg] \, .
 \label{boxf} 
\end{align}
Note that the combination proportional to the IR factor $\left (- \frac{1}{\epsilon_{\mathrm{IR}}}+\ln\frac{\mu^2}{\mu_{\mathrm{IR}}^2} \right )$ in the second line of Eq. (\ref{boxf})  cancels the IR part of the bremsstrahlung 
contribution (\ref{brems}).

\subsection{Other vertex  diagrams} 

The remaining vertex diagrams (1b), ($1{\rm b}'$) , (1c), (${\rm 1e}$), (${\rm 1e}'$)  and (1f) vanish in Feynman gauge. 
In particular, 
for the diagram shown in Fig.~(1b), we have
\begin{align}
{g^2 C_A\cdot}	6i \varepsilon_{\alpha\mu\nu\rho}V_{110}^{\alpha}p^{\mu}\epsilon_1^{\nu}\epsilon_2^{*\rho}=0 \  , 
\end{align}
since $V_{110}^{\alpha} \sim p^\alpha$ according to Eq. (\ref{Vmu110}).
For the diagram ($1{\rm b}'$),  the result is 
\begin{align}
{g^2 C_A\cdot }	6i \varepsilon_{\alpha\mu\nu\rho} \epsilon_1^{\nu}  \epsilon_2^{*\rho} (p^{\mu}V_{011}^{\alpha}
	+q^{\mu} V_{011}^{\alpha}-p^{\alpha}q^{\mu} S_{011}) \  . 
\end{align}
It also  vanishes  after we 
 use $V_{011}^{\alpha} =-((1-x)q^{\alpha}-x p^{\alpha}) S_{011}$ (see Eqs. (\ref{S011}), (\ref{Vmu011})). 
For the diagram (1c), the result is identically zero.  The contributions of the  diagrams
(${\rm 1e}$) 
\begin{align}
{g^2 C_A}	\frac{2i}{n\cdot P(1-x)}\varepsilon_{\alpha\beta\mu\nu}n^{\alpha}p^{\beta}\epsilon_1^{\mu}\epsilon_2^{*\nu}S_{010}
\end{align}
and (${\rm 1e}'$)
\begin{align}
{g^2 C_A}	\frac{2i}{n\cdot P(1-x)}\varepsilon_{\alpha\beta\mu\nu}n^{\alpha}(p+q)^{\beta}\epsilon_1^{\mu}\epsilon_2^{*\nu}S_{010} 
\end{align}
 are proportional to the function 
\begin{align}
	&S_{010}
	=\int\frac{\dd ^dk}{(2\pi)^d k^2} \delta \left (x-1-\frac{n\cdot k}{n\cdot P} \right ) 	\ . 
	\end{align}
	Substituting the  integrand factor by 
	\begin{align}
 \frac1{k^2} \delta \left (x-1-\frac{n\cdot k}{n\cdot P} \right ) = \frac1{2\pi i}
 \int_{-\infty}^\infty \dd \gamma \, e^{i \gamma(x-1-{n\cdot k}/{n\cdot P} ) }
 \int_0^\infty \dd \alpha e^{i\alpha k^2} 
	 	\ , 
	\end{align}
	and using the fact  that   the resulting Gaussian $k$-integral does not depend on $n$ when  $n^2=0$, 
		\begin{align}
\int   \dd ^dk e^{i\alpha k^2 - i \gamma {n\cdot k}/{n\cdot P} }= 
\int   \dd ^dk e^{i\alpha k^2 }	\ , 
	\end{align}
	we see that  
 $S_{010}$  reduces to 
\begin{align}
	&S_{010}
	 = 
	 \delta (x-1) \int\frac{\dd ^dk}{(2\pi)^d k^2} 
	\ ,
	\end{align}
i.e., to  the integral containing just one propagator.   Such integrals are treated as zero in the dimensional regularization.

Finally, the contributions of the four-gluon vertex  diagram 
(1f) is identically zero.

\subsection{Self-energy-type   diagrams} 

We should also  include the contributions of the diagrams of self-energy type. 
They have both UV and IR logarithmic divergences.
We will present here the results for $x>0$, understanding that one should 
complement them by the  $\{x \to -x\}$ contributions in the final result.  In particular, 
the diagram (1g) is given by 
\begin{align}
	&-i \delta(1-x) \varepsilon(p,q,\epsilon_1,\epsilon_2^*) g^2 C_A  \int\frac{\dd ^dk}{(2\pi)^d}\frac{3}{k^2 (p-k)^2}
	\end{align}
	which produces 
\begin{align}	
& \delta(1-x) \varepsilon(p,q,\epsilon_1,\epsilon_2^*)  \frac{\alpha_s C_A}{\pi}\frac34 
\left (\frac{1}{\epsilon}-\frac{1}{\epsilon_{\mathrm{IR}}}+\ln\frac{\mu^2}{\mu_{\mathrm{IR}}^2}
\right )
\end{align}
in the \msbar  scheme. 
Its mirror-conjugate diagram 
$({\rm 1g}')$ gives the same contribution
\begin{align}
	& \delta(1-x)  \varepsilon(p,q,\epsilon_1,\epsilon_2^*)  \frac{\alpha_s C_A}{\pi}\frac34
	\left (\frac{1}{\epsilon}-\frac{1}{\epsilon_{\mathrm{IR}}}+\ln\frac{\mu^2}{\mu_{\mathrm{IR}}^2} \right ) \ . 
\end{align}
The self-energy corrections (${\rm 1h}$),  (${\rm 1h}'$) to   the external gluon lines produce 
\begin{align}
	Z_g \tilde{F}^{(0)}(x)=-2\varepsilon(p,q,\epsilon_1,\epsilon_2^*) \delta(1-x)  \frac{\alpha_s}{\pi}
	\left (\frac{5}{12}C_A-\frac13 T_F n_f \right )
	\left (\frac{1}{\epsilon}-\frac{1}{\epsilon_{\mathrm{IR}}}+\ln\frac{\mu^2}{\mu_{\mathrm{IR}}^2}  \right  ) 
	\ .
\end{align}

\section{One-loop corrections in light-cone gauge}
To verify  gauge invariance of our results presented  in the previous  section, we also perform a calculation 
in the light-cone gauge.  The well-known advantage of the light-cone gauge is the absence of the gauge link. 
As a result,   the diagrams 
\ref{real}(b), \ref{real}(b$'$),  \ref{real}(e), \ref{real}(e$'$)  are automatically zero, and ``everything'' comes from the box diagram 
~\ref{real}(a), for which   we have obtained 
	\begin{align}
		&	\widetilde{F}_{(1a)}(x,p,q,\epsilon_1,\epsilon_2)\bigg|_{x\geq 0}\nonumber\\
		&=g^2 C_A\cdot 2i\varepsilon_{\alpha\mu\nu\rho}\bigg[\bigg(\frac{2}{1-x} V_{011}^{\alpha}+\frac{2}{1-x} V_{110}^{\alpha}-2q^2 V_{111}^{\alpha}-V_{101}^{\alpha} \bigg ) q^{\mu}\epsilon_1^{\nu}\epsilon_2^{*\rho}\nonumber\\
		&+ \frac{2}{x} \frac{1}{n\cdot P} V_{101}^{\alpha} n^{\mu}q^{\nu}\epsilon_2^{*\rho}q\cdot \epsilon_1 - \left (\frac{1}{x}V_{101}^{\alpha}+\frac{2}{1-x}V_{011}^{\alpha} \right ) \frac{1}{n\cdot P} n^{\mu}q^{\nu}\epsilon_1^{\rho}q\cdot \epsilon_2^*\nonumber\\
		&+4 V_{111}^{\alpha} p^{\mu} q^{\nu} \left (\epsilon_2^{*\rho}q\cdot \epsilon_1 -\epsilon_1^{\rho} q\cdot \epsilon_2^*
		\right )\nonumber\\
		&+\frac{2}{1-x}\frac{1}{n\cdot P}  n^{\mu}q^{\nu} \left (T_{110}^{\alpha\beta}\epsilon_2^{*\rho}\epsilon_{1,\beta}-T_{011}^{\alpha\beta}\epsilon_1^{\rho}\epsilon_{2,\beta}^* \right ) - \frac{3}{x}\frac{1}{n\cdot P} T_{101}^{\alpha\beta} n^{\mu}q^{\nu}
		\left (\epsilon_1^{\rho}\epsilon_{2,\beta}^*-\epsilon_{2}^{*\rho}\epsilon_{1,\beta} \right )\nonumber\\
		& +8 T_{111}^{\alpha\beta}  p^{\mu} q^{\nu}
		\left  (\epsilon_2^{*                          \rho}\epsilon_{1\beta}-\epsilon_1^{\rho}\epsilon_{2,\beta}^* \right ) \bigg]\, .
	\end{align}
	Using the expressions for  the integrals given in  the Appendix, we get 
	\begin{align}
		&\delta(x)\frac{g^2 C_A q^2}{96\pi^2 n\cdot P x(1-x)}\varepsilon(n,q,\epsilon_1,\epsilon_2^*)\bigg[(6-17x(1-x))
		\left (\frac{1}{\epsilon}+\ln\frac{\mu^2}{-q^2} \right )+\frac{16 \left (3-7x(1-x) \right )}{3}\bigg]\nonumber\\
		&+\delta'(x)\frac{g^2 C_A q^2}{96\pi^2 n\cdot P (1-x)}\varepsilon(n,q,\epsilon_1,\epsilon_2^*)\bigg(\frac{1}{\epsilon}+\ln\frac{\mu^2}{-q^2}+\frac83\bigg)\nonumber\\
		&+\theta(0\leq x< 1)\bigg[-\frac{g^2 C_A}{2\pi^2}\varepsilon(p,q,\epsilon_1,\epsilon_2^*)
		\left (\frac{2-3x+2x^2}{1-x} \left (\frac{1}{\epsilon}+\ln\frac{\mu^2}{-q^2(1-x)^2}\right )+4(1-x) \right )
		\nonumber\\
		&+\frac{g^2 C_A q^2}{8\pi^2 n\cdot P}\varepsilon(n,q,\epsilon_1,\epsilon_2^*)
		\left (\frac{1}{\epsilon}+\ln\frac{\mu^2}{-q^2(1-x)^2}\right )\nonumber\\
		&+\frac{g^2 C_A}{4\pi^2 n\cdot P}\left (\varepsilon(n,p,q,\epsilon_1)q\cdot \epsilon_2^*+\varepsilon(n,p,q,\epsilon_2^*)
		q\cdot \epsilon_1\right  )
		\nonumber\\
		& \times  \left ((1-2x) 
		\left (\frac{1}{\epsilon}+\ln\frac{\mu^2}{-q^2(1-x)^2}\right )+2(1-x)\right )\bigg]\, ,
	\end{align}	
where we have expanded the results in $\epsilon$.
	Now,  recalling Eqs.~\eqref{eq:str1} and \eqref{eq:str2},  we find that 
	 only the structure $\varepsilon(p,q,\epsilon_1,\epsilon_2^*)$ is left, and we have  
	\begin{align}
&	\widetilde{F}_{(1a)}(x,p,q,\epsilon_1,\epsilon_2)\bigg|_{x\geq 0}\nonumber\\
		=	&\frac{\alpha_s C_A}{\pi}\Pi(p,q,\epsilon_1,\epsilon_2^*)\theta(0\leq x< 1)\bigg[\frac{1}{1-x}\bigg(\frac{1}{\epsilon}+\ln\frac{\mu^2}{-q^2(1-x)^2}\bigg)+2(1-x)\bigg]+\{x\to -x\}\, ,
	\end{align}
where  the $-1<x\leq 0$ part is also included. One can observe that this expression 
 is the same as the $x\neq \pm 1$ part of the  total  result  obtained in Feynman gauge.

There are  four other  diagrams \ref{real}(b), \ref{real}(b$'$), \ref{real}(c), \ref{real}(f),   which, in principle,  could contribute corrections for $x\neq \pm 1$ in
the  light-cone gauge.  However, they produce no such contributions. To begin with,
one can find that \ref{real}(c) is zero by contracting the indices.  
For Fig.~\ref{real}(b), we have
\begin{align}
	g^2 C_A \bigg(\frac{1}{x}V_{001}^{\alpha}+\frac{1}{1-x}V_{010}^{\alpha}\bigg)\frac{2i}{n\cdot P}\varepsilon_{\alpha\mu\nu\rho}n^{\mu}\epsilon_1^{\nu}\epsilon_2^{*\rho}  \,  , 
\end{align}
and for Fig.~\ref{real}(b$'$) we obtain 
\begin{align}
	g^2 C_A \left (\frac{1}{1-x} V_{010}^{\alpha}+\frac{1}{x}V_{100}^{\alpha} \right )\frac{2i}{n\cdot P}\varepsilon_{\alpha\mu\nu\rho}n^{\mu}\epsilon_1^{\nu}\epsilon_2^{*\rho} \ . 
\end{align}
For Fig.~\ref{real}(f), we have
\begin{align}
-	g^2 C_A  \frac{1}{2x}V_{101}^{\alpha}n^{\mu}q^{\nu}(\epsilon_1^{\rho}q\cdot \epsilon_2^*+\epsilon_2^{*\rho}q\cdot \epsilon_1)\frac{2i}{n\cdot P}\varepsilon_{\alpha\mu\nu\rho}\, .
\end{align}
According to the integrals listed in the Appendix, 
the contributions from the diagrams \ref{real}(b), \ref{real}(b$'$) and \ref{real}(f) are all zero. 
Hence, the only nonzero  correction for  $x\neq \pm 1$  is from the box diagram Fig.~\ref{real}(a), and we get the same result
 for  $x\neq \pm 1$  in both Feynman and light-cone gauges.

The calculation of the $\sim \delta (1\pm x)$ 
contributions  produced  in the light-cone gauge by self-energy-type diagrams 
 is a well-known routine, and we skip it.  
Some of such   contributions  combine with the $x\neq \pm 1$ terms into the ``plus-prescription'' expressions, and the others 
generate  anomalous 
dimensions of the local operator.

\section{Total result }

Combining the contributions from all diagrams, we get  the total result, which,
 including the tree-level contribution,   reads
\begin{align}
F(x,q^2; \mu^2, \mu_{\mathrm{IR}}^2)
=& \Pi(p,q,\epsilon_1,\epsilon_2^*) 
\nn & \times \Bigg \{ 1
 +
\frac{\alpha_s}{\pi }C_A \bigg\{{\theta(0\leq x\leq 1)}\bigg[\frac{1}{1-x}\bigg(\frac{1}{\epsilon}+\ln\frac{\mu^2}{-q^2(1-x)^2}\bigg)+2(1-x)\bigg]\bigg\}_+\nonumber\\
&\hspace{1cm} +\frac{\alpha_s}{4\pi}\beta_0\delta(1-x)
\left (\frac{1}{\epsilon}-\frac{1}{\epsilon_{\mathrm{IR}}}+\ln\frac{\mu^2}{\mu_{\mathrm{IR}}^2} \right )
\nonumber\\
&\hspace{1cm}  -\frac{\alpha_s}{\pi}C_A \delta(1-x)\bigg(\frac{1}{2\epsilon_{\mathrm{IR}}^2}+\frac{1}{2\epsilon_{\mathrm{IR}}}\ln\frac{\mu_{\mathrm{IR}}^2}{-q^2}+\frac14 \ln^2\frac{\mu_{\mathrm{IR}}^2}{-q^2}-\frac{\pi^2}{24}-1\bigg) \Bigg \} \nn & 
\hspace{1cm}  +\{x\to -x\}  \  . 
\label{total} 
\end{align}
The coefficient accompanying the $1/\epsilon$  pole (multiplied by  the $\alpha_s /2 \pi$ factor) 
gives  the evolution kernel 
   \begin{align}
 P_{gg}^{\widetilde F}(x) = 
 \frac{\beta_0}{2} \delta (1-x) + C_A   \bigg [\frac{2}{1-x}\theta(0\leq x \leq 1)\bigg ]_+
 +\{x\to -x\}  
\  
\end{align}
for $\widetilde F (x,q^2)$. 
It  has two ingredients. The $\sim \beta_0$ term corresponds to the anomalous  dimension of the 
local operator $F^{\mu \nu}(0) \widetilde{F}_{\mu \nu}(0)$. 
The ``plus-prescription'' term, displayed in the second line of Eq. (\ref{total}),
  is specific for the nonlocal case.  
Note that it  does not contain the IR poles ${\epsilon_{\mathrm{IR}}}$
and the IR scale $\mu_{\mathrm{IR}}$.  
As already mentioned, the  terms  
$\sim \left 
(- {1}/{\epsilon_{\mathrm{IR}}}+\ln {\mu^2}/{\mu_{\mathrm{IR}}^2} \right )$ 
present in the box and bremsstrahlung diagrams,
cancel each other.
As a result,  the IR cutoff in this term 
is provided by  the momentum transfer $q^2$, just like in the case of  the 
 ``gluon condensate'' PDF $F(x)$ discussed in 
our recent paper \cite{Radyushkin:2021fel}
(a similar observation was made in the studies of the quark GPDs \cite{Liu:2019urm}, \cite{Ji:2015qla}).

Another observation is that the kernel $ P_{gg}^{\widetilde F}(x) $ 
 coincides with that for   the  ``gluon condensate'' PDF $F(x)$, 
 despite the difference in the structure of the relevant nonlocal operators.   
However, our expression for the evolution kernel does not coincide with  that obtained in Ref. \cite{Hatta:2020ltd}.

The UV 
finite ``Sudakov''  term,  shown in the $4{\rm th}$ line  of  Eq. (\ref{total}),   is 
an artifact of the IR regularization by a finite momentum transfer $q$.
Recall that,  to maintain  the necessary strict gauge invariance in our calculations, 
we have chosen  to  take    zero-virtuality initial and final momenta $p_1,p_2$.
Next, to get a non-vanishing result for the overall kinematical factor   $\Pi (p,q,\epsilon_{1}, \epsilon_{2}^* )$  
 (see Eq. (\ref{F0})), we
have  imposed 
 a  nonzero momentum transfer $q=p_2-p_1$, with $q^2 \neq 0$. 
 As a result, 
 the  box diagram (1a)   is formally in the Sudakov kinematics $-q^2 \gg |p_1^2|\sim |p_2^2|$,
  which is signalized by double 
  logarithms  in the Sudakov term.  
  Because of its purely IR nature, we  may  absorb the ``Sudakov'' term into a ``bare'' GPD. 
In other words, since it does not contain the  UV parameter  
$\mu$, the  ``Sudakov'' term does not affect 
the relation between the functions $\widetilde F^{(1)} (x,q^2; \mu^2)$ at different 
evolution scales $\mu$.  Similarly, calculating the
matrix element $\langle p_2 | F^{\mu \nu} \left (-{z}/{2} \right ) 
W \left [-{z}/{2}, {z}/{2} \right ] \widetilde F_{\mu \nu}({z}/{2} | p_1 \rangle $ for $z^2 \neq 0$ (i.e., off the light cone,
which is necessary for lattice calculations of $\widetilde F(x,q^2; \mu^2)$),
one would get the same  Sudakov terms, that would cancel in the matching condition 
between off-the-light-cone and  on-the-light-cone versions of the GPD.

Finally, we would like to  mention that we do not have  $\delta(x)$ terms in our one-loop result (\ref{total}) 
 which one could  identify as a ``zero-mode'' contribution.


   \section{Summary and outlook}

In this paper, we have presented the results of   one-loop corrections in the 2-gluon sector  to 
the ``topological charge'' GPD $\widetilde F(x,q^2)$ introduced in Ref. \cite{Hatta:2020ltd}.
Just like in our \mbox{paper \cite{Radyushkin:2021fel} } about the ``gluon condensate'' PDF $F(x)$,
to get a nonzero contribution for the gluon matrix element at the tree level 
and maintain gauge invariance, 
we took 
a nonforward matrix element between on-shell massless gluons, i.e. we  have considered a  GPD
 reducing to $\widetilde F(x)$ in the forward limit.
Ref. \cite{Hatta:2020ltd} also deals with a GPD,  
 however, the calculation was done 
 for off-shell gluons, which violates  gauge invariance. 
 
 We have performed our calculations with on-shell external gluons both in
 Feynman and light-cone gauges, and obtained the same result. 
Our  Feynman-gauge and light-cone-gauge calculations are   described   in the present 
 paper on the diagram-by-diagram level. They  give  a result differing from that of Ref.  \cite{Hatta:2020ltd},
 thus demonstrating once more the importance of  doing the calculations of gluon matrix elements   in a
 strict compliance with the gauge-invariance requirements.  

In Ref.  \cite{Ji:2020baz}, it was suggested that some twist-4 gluon PDFs may have 
$\delta (x)$ zero-modes, similar to those observed 
 in one-loop perturbative QCD expressions for the 
 twist-3 quark PDFs (see, e.g.,  \cite{Burkardt:2001iy}).  
 However, our one-loop expression  for the twist-4 gluon 
 GPD  $\widetilde F(x,q^2)$ does not contain such terms. 
 
 It should {be} emphasized that our calculation deals with the matrix elements of the  \mbox{twist-4 }
 bilocal operator 
 $F^{\mu \nu} \left (-{z}/{2} \right ) 
 \widetilde F_{\mu \nu}({z}/{2}) $  (we skip the  link factor $W$ here and below) 
between   two external gluon states.  
In the OPE language, this means that  we are picking out  the $F^{\mu \nu} \left (u z\right ) 
 \widetilde F_{\mu \nu}(v z)  $  terms in the expansion of the original operator 
 product $F^{\mu \nu} \left (-\frac{z}{2} \right ) 
 \widetilde F_{\mu \nu}(\frac{z}{2} )$. 
However, one can easily imagine  twist-4 nonlocal operators built from three  and even four gluon fields
(like 
$z^\alpha z^\beta F_{\alpha \mu} \left (u z \right ) F^{ \mu \nu} (v z)
 \widetilde F_{\nu\beta}(wz)$, etc.).  To pick out coefficient functions corresponding to such operators,
 one should consider matrix elements of $F^{\mu \nu} \left (-{z}/{2} \right ) 
 \widetilde F_{\mu \nu}({z}/{2} )$ between three  and four external gluons.   
 In the momentum representation, such a procedure of calculating  the mixing between different types 
 of gluon operators  involves some element of guessing and uncertainty about whether 
all possible combinations have been taken into account.

Another  way   to approach this problem is to 
calculate  corrections in the operator form, without  making projections
on external states  at all, like it was done  in Refs. \cite{Radyushkin:2017lvu,Balitsky:2019krf,Balitsky:2021qsr,Balitsky:2021cwr}  for
the ``twist-2''  
quark   and   gluon bilocal operators 
outside the   light cone. This gives   a possible direction  for future 
studies of the twist-4 gluon distributions. 
A natural first step would be a coordinate-space formulation of the results obtained 
using the momentum-space techniques 
 in  the present paper and in Ref. \cite{Radyushkin:2021fel}.

\section*{ Acknowledgements}

We thank I. Balitsky and W. Morris for discussions and Y. Hatta for correspondence. 
This work is supported by Jefferson Science Associates,
 LLC under  U.S. DOE Contract  \mbox{\#DE-AC05-06OR23177} 
 and by U.S. DOE Grant \#DE-FG02-97ER41028.


\appendix

\section{Table of integrals}

\begin{align}
\label{S010} 
	S_{010}\sim & 0,\\
	\label{S011} 
	S_{011}=&\frac{i}{16\pi^2}\bigg(\frac{1}{\epsilon}-\frac{1}{\epsilon_{\mathrm{IR}}}+\ln\frac{\mu^2}{\mu_{\mathrm{IR}}^2}\bigg){\theta(0\leq x\leq 1)},\\
	\label{Vmu100}
	V_{100}^{\alpha}\sim & V_{010}^{\alpha}\sim V_{001}^{\alpha}\sim 0,\\
	\label{Vmu110} 
	V^{\mu}_{110}
	{\sim}&\frac{i}{16\pi^2}\bigg(\frac{1}{\epsilon}-\frac{1}{\epsilon_{\mathrm{IR}}}+\ln\frac{\mu^2}{\mu_{\mathrm{IR}}^2}\bigg)x p^{\mu}{\theta(0\leq x\leq 1)},\\
	\label{Vmu011} 
	V^{\mu}_{011}{\sim}&-\frac{i}{16\pi^2}\left(\frac{1}{\epsilon}-\frac{1}{\epsilon_{\mathrm{IR}}}+\ln\frac{\mu^2}{\mu_{\mathrm{IR}}^2}\right)((1-x)q^{\mu}-x p^{\mu}){\theta(0\leq x\leq 1)},\\
	\label{Vmu101} 
	V^{\mu}_{101}
	=&-\frac{i}{16\pi^2} \left (\frac{\mu^2 e^{\gamma_E}}{-q^2} \right )^{\epsilon} 
	\Bigg  [\Gamma(\epsilon) \frac{\Gamma(2-\epsilon)\Gamma(1-\epsilon)}{\Gamma(3-2\epsilon)}q^{\mu}\delta(x)
	 +\frac{1}{2}\frac{q^2}{ n\cdot P }\frac{\Gamma(2-\epsilon)^2}{\Gamma(4-2\epsilon)}\Gamma(-1+\epsilon) n^{\mu}\delta'(x)
	\Bigg  ],\\
	\label{Vmu111} 
	V^{\mu}_{111}
	= &- \frac{i}{16\pi^2 q^2}  (\frac{\mu_{\mathrm{IR}}^2 e^{\gamma_E}}{-q^2})^{\epsilon_{\mathrm{IR}}}\Gamma(1+\epsilon_{\mathrm{IR}})\frac{\Gamma(-\epsilon_{\mathrm{IR}})^2}{\Gamma(-2\epsilon_{\mathrm{IR}})} \bigg[\frac12(1-x)^{-2\epsilon_{\mathrm{IR}}} q^{\mu}\nonumber\\
&- x(1-x)^{-1-2\epsilon_{\mathrm{IR}}} p^{\mu}\bigg] {\theta(0\leq x\leq 1)}\nonumber\\
&+\frac{i}{32\pi^2}\bigg(\frac{\mu^2 e^{\gamma_E}}{-q^2}\bigg)^{\epsilon}\frac{\Gamma(1-\epsilon)^2 \Gamma(\epsilon)}{\Gamma(2-2\epsilon)}\frac{n^{\mu}}{n\cdot P}\bigg[-\delta(x)\nonumber\\
&+(1-2\epsilon)(1-x)^{-2\epsilon}{\theta(0\leq x\leq 1)}\bigg],\\ 
	\label{Tmunu110} 
	T^{\mu\nu}_{110}\sim&\frac{i}{16\pi^2}\bigg(\frac{1}{\epsilon}-\frac{1}{\epsilon_{\mathrm{IR}}}+\ln\frac{\mu^2}{\mu_{\mathrm{IR}}^2}\bigg)x^2 p^{\mu}p^{\nu}{\theta(0\leq x\leq 1)},\\
	\label{Tmunu011} 
	T^{\mu\nu}_{011}\sim&\frac{i}{16\pi^2}\bigg(\frac{1}{\epsilon}-\frac{1}{\epsilon_{\mathrm{IR}}}+\ln\frac{\mu^2}{\mu_{\mathrm{IR}}^2}\bigg)[(1-x)q^{\mu}-x p^{\mu}][(1-x)q^{\nu}-x p^{\nu}]{\theta(0\leq x\leq 1)},\\
	\label{Tmunu101}
	T^{\mu\nu}_{101}
\sim & \frac{i}{16\pi^2}(\frac{\mu^2 e^{\gamma_E}}{-q^2})^{\epsilon}\bigg(q^2 \frac{g^{\mu\nu}}{2}\frac{\Gamma^2(2-\epsilon)}{\Gamma(4-2\epsilon)}\Gamma(-1+\epsilon)+\frac{\Gamma(3-\epsilon)\Gamma(1-\epsilon)}{\Gamma(4-2\epsilon)}\Gamma(\epsilon)q^{\mu}q^{\nu}\bigg)\delta(x)\nonumber\\
&+\frac{i}{16\pi^2}(\frac{\mu^2 e^{\gamma_E}}{-q^2})^{\epsilon}\frac{q^2}{2}\frac{\Gamma^2(2-\epsilon)}{\Gamma(4-2\epsilon)}\Gamma(-1+\epsilon)\frac{q^{\mu}n^{\nu}+q^{\nu}n^{\mu}}{n\cdot P}\delta'(x),\\	
		\label{Tmunu111} 
	T^{\mu\nu}_{111}
{\sim}& \frac{i}{16\pi^2} \bigg\{\frac{g^{\mu\nu}}{2} (1-x)^{1-2\epsilon}\frac{\Gamma(1-\epsilon)^2}{\Gamma(2-2\epsilon)}\Gamma(\epsilon)(\frac{\mu^2 e^{\gamma_E}}{-q^2})^{\epsilon} \nonumber\\
&+\frac{1}{q^2}\Gamma(1+\epsilon_{\mathrm{IR}})(\frac{\mu_{\mathrm{IR}}^2 e^{\gamma_E}}{-q^2})^{\epsilon_{\mathrm{IR}}} \bigg[(1-x)^{1-2\epsilon_{\mathrm{IR}}}\frac{\Gamma(2-\epsilon_{\mathrm{IR}})\Gamma(-\epsilon_{\mathrm{IR}})}{\Gamma(2-2\epsilon_{\mathrm{IR}})} q^{\mu}q^{\nu}\nonumber\\
&-x (1-x)^{-2\epsilon_{\mathrm{IR}}}\frac{\Gamma(1-\epsilon_{\mathrm{IR}})\Gamma(-\epsilon_{\mathrm{IR}})}{\Gamma(1-2\epsilon_{\mathrm{IR}})}(q^{\mu}p^{\nu}+p^{\mu}q^{\nu})\nonumber\\
&+x^2(1-x)^{-1-2\epsilon_{\mathrm{IR}}}\frac{\Gamma(-\epsilon_{\mathrm{IR}})^2}{\Gamma(-2\epsilon_{\mathrm{IR}})} p^{\mu}p^{\nu}\bigg]\bigg\} {\theta(0\leq x\leq 1)}\nonumber\\
&-\frac{i}{32\pi^2 n\cdot P}\bigg(\frac{\mu^2 e^{\gamma_E}}{-q^2}\bigg)^{\epsilon}\Gamma(\epsilon)\frac{\Gamma^2(1-\epsilon)}{\Gamma(2-2\epsilon)}\bigg\{\bigg[-\frac12\delta(x)\nonumber\\
&+(1-\epsilon)(1-x)^{1-2\epsilon}{\theta(0\leq x\leq 1)}\bigg](n^{\mu}q^{\nu}+n^{\nu}q^{\mu})\nonumber\\
&+(1-x)^{-2\epsilon}(1-2(1-\epsilon)x)(n^{\mu}p^{\nu}+n^{\nu}p^{\mu}){\theta(0\leq x\leq 1)}\bigg\}.
\end{align}

{\it Note:} The sign $\sim$ means that, in addition to the explicitly written terms, the  contribution 
also contains $\int \dd ^dk/k^2$ terms which are treated as zero. {The terms that vanish under contraction of indices are also neglected.}

 \bibliography{GT4.bib}
\bibliographystyle{jhep}
 
 \end{document}